\documentclass[useAMS,usenatbib]{mn2e}

\usepackage{hyperref}
\usepackage[utf8]{inputenc}
\usepackage[T1]{fontenc}
\usepackage{amsmath}
\usepackage{amssymb}
\usepackage{cancel}
\usepackage{units}
\usepackage{array}
\usepackage{graphicx}

\usepackage{color}

\newcommand{\apj}{ApJ}
\newcommand{\apjl}{ApJ}
\newcommand{\mnras}{MNRAS}
\newcommand{\aap}{A\&A}

\newcommand{\nat}{Nature}

\title[Mixing of metals]{Mixing and transport of metals by gravitational instability-driven turbulence in galactic discs}
\author[A.~Petit et al.]
{Antoine~C.~Petit,$^{1,2}$ Mark~R.~Krumholz,$^2$, Nathan J.~Goldbaum$^2$, \& John C.~Forbes$^2$\\\
	$^1$ICFP undergraduate program, Physics department,  \'Ecole Normale Sup\'erieure, 45 Rue d'Ulm 75005, Paris, France\\
	$^2$Department of Astronomy and Astrophysics, University of California, 1156 High Street, Santa Cruz, CA 95064, USA}
\date{\today}

\pagerange{\pageref{firstpage}--\pageref{lastpage}} \pubyear{2014}

\newcommand{\mixtime}{100\  Myr}
\newcommand{\mixtimem}{100\mbox{ Myr}}

\renewcommand\P{\mathcal{P}}

\renewcommand{\d}{\mathrm{d}}

\newcommand{\red}[1]{\textcolor{black}{#1}}

\begin{document}

\label{firstpage}

\maketitle
\begin{abstract}
Metal production in galaxies traces star formation, and is highly concentrated toward the centers of galactic discs. This suggests that galaxies should have inhomogeneous metal distributions with strong radial gradients, but observations of present-day galaxies show only shallow gradients with little azimuthal variation, implying the existence of a redistribution mechanism. We study the role of gravitational instability-driven turbulence as a mixing mechanism by simulating an isolated galactic disc at high resolution, including metal fields treated as passive scalars. Since any cylindrical field can be decomposed into a sum of Fourier-Bessel basis functions, we set up initial metal fields characterized by these functions and study how different modes mix. We find both shear and turbulence contribute to mixing, but the mixing strongly depends on the symmetries of the mode. Non-axisymmetric modes have decay times smaller than the galactic orbital period because shear winds them up to small spatial scales, where they are erased by turbulence. The decay timescales for axisymmetric modes are much greater, though for all but the largest-scale inhomogeneities the mixing timescale is still short enough to erase chemical inhomogeneities over cosmological times. These different timescales provide an explanation for why galaxies retain metallicity gradients while there is almost no variation at a fixed radius. Moreover, the comparatively long timescales required for mixing axisymmetric modes may explain the greater diversity of metallicity gradients observed in high redshift galaxies as compared to local ones: these systems have not yet reached equilibrium between metal production and diffusion.
\end{abstract}

\section{Introduction}

	Understanding the dynamics of the  flow of metals through and around galaxies is a key problem in the study of galaxy formation. Metals  trace  the history of the gas flows in galaxies such as the young Milky Way \citep{Ivezic2012} and record the buildup of stellar populations in their progenitors at high redshift \citep{Tremonti2004,Erb2006}. Moreover, metals are not just passive tracers. They change the chemical and radiative cooling properties of the interstellar medium, altering how stars form  \citep{Krumholz2009,Krumholz2011,Krumholz2012a, Krumholz2012b, Krumholz2013a, Glover2012a, Glover2012b}.
	
	It is possible to observe the spatial distribution of metals in the Milky Way \citep{Henry2010,Balser2011,Luck2011,Yong2012}, in nearby galaxies \citep{Vila-Costas1992,Considere2000,Pilyugin2004,Kennicutt2011}, and even in the high redshift universe \citep{Cresci2010,Jones2013}.  The distribution differs from one galaxy to another in the local universe, but generally, a radial gradient of the order of $-0.03$ dex kpc$^{-1}$ is observed; the negative sign means that metallicity decreases at large radii. High redshift galaxies are far less regular, and show patterns that vary from flat or slightly positive gradients to some even steeper than in the local universe.
	
	Metal production in a galaxy is continuously fed by the stellar feedback \citep{Phillipps1991}. Without large-scale mixing, the density of metals should be directly proportional to the stellar density. However it has been observed that outer galaxy regions have a metallicity that is significantly higher than one would expect if the only metals present were those produced by stars at the same galactocentric radius, while the inner regions of galaxies have metallicities much smaller than one would expect if all locally-produced metals remained part of the disc \citep{Bresolin2009,Bresolin2012,Werk2011}. Moreover, galaxies show \red{gas-phase} metallicity inhomogeneities of at most a few tenths of a dex at fixed galactocentric radius \citep{Rosolowsky2008, Bresolin2011, Sanders2012, Berg2013}, while star formation is far patchier. All of these observations imply that metal transport must play an important role in determining the metallicity distribution of galaxies.

	Unfortunately, metal transport theories remain very primitive.  The main problem  is that it is very difficult to run high resolution simulations of entire galaxies from redshift $z\simeq 1-2$, when the bulk of the stars in present-day discs formed, to the present-day. Simulations that do include explicit metal tracking over cosmological times are generally forced by resolution limits to adopt a temperature floor of $\sim 10^4$ K, and are thus unable to provide a realistic treatment of transport through the multi-phase interstellar medium (ISM) of a galactic disc \citep[e.g.][]{Wiersma2009, Pilkington2012a, Pilkington2012b, Brook2012,  Few2012a, Few2012b, Minchev2013,  Minchev2014}. The alternative is parameterized 1D models \citep[e.g.][]{Chiappini2001, Spitoni2011, Forbes2013}, but for these cases the rate of metal transport is a free parameter, not a prediction of the model.
	
	In order to study mixing and transport within galactic discs, the only solution is to do local studies. \red{A large number of such studies have been published focusing on the radial migration of stars through a galactic disc \citep[e.g.,][to name only a few]{brunetti11a, bird12a, di-matteo13a, grand14a}, but much less work has been done focusing on the gas.} \citet{de-Avillez2002} performed early studies of supernova-driven turbulent mixing, and more recently \citet{Yang2012} used shearing-box simulations to show that turbulence driven by thermal instability is very efficient at mixing metals. \red{They also demonstrated that the multi-phase nature of the ISM has dramatic consequences for the mixing of metals, implying that only simulations with enough resolution to allow such a multi-phase medium to form are likely to produce reliable results.} However, both \red{\citeauthor{de-Avillez2002}'s and \citeauthor{Yang2012}'s} studies were limited to small portions of a galaxy, making it impossible to study transport at the galactic scale, driven by geometrical effects and galaxy-scale processes like spiral arms. \red{\citet{kubryk13a} simulated an entire isolated disc galaxy to measure the effects of a bar on radial mixing of gas and stars, and \citet{kubryk15b} used the results of this simulation to construct a semi-analytic model of element mixing. However, these simulations suffered from much the same resolution limitations as the cosmological ones, in that they used a $\sim 10^4$ K temperature floor and thus lacked a multi-phase ISM. The highest-resolution study performed to date is that of \citet{grand15a}, who studied stellar and gas migration in an isolated galaxy simulated at $\sim 150$ pc resolution, which is still not sufficient to capture the cold phase of the atomic ISM.}
	
	Our goal in this paper is to study the transport of metals in a global simulation of a large spiral galaxy, including the effects of spiral structure and thermal and gravitational instability. To this end, we simulate an isolated galaxy at $20$ pc resolution, \red{roughly an order of magnitude higher in resolution than previous studies,} and with a cooling floor that is low enough to allow development of a full multi-phase atomic medium. In Section \ref{setup} we describe the setup of our simulations, and how we treat the metal fields. Section \ref{analysis} presents a quantitative analysis of mixing. In Section \ref{implications} we discuss the astrophysical implications of our work. Finally, we summarize in Section \ref{conclusion}.

\section{Simulation setup}
\label{setup}

	\subsection{The isolated galaxy}
		
	To study the turbulent mixing, we simulate an isolated galaxy using the Adaptive Mesh Refinement (AMR) code ENZO \citep{Bryan2014}. ENZO includes fluid dynamics, gravity, sink particles, and radiative cooling implemented with the cooling and chemistry library Grackle\footnote{https://grackle.readthedocs.org/}. We do not include magnetic fields since \citet{Yang2012} have shown that they have a very a small effect on the turbulent mixing on galactic scales. We model star formation such that, when the gas density in any cell reaches a threshold density of 50 particles per cubic centimeter, there is a probability equal to $\epsilon_{\mathrm{SF}}m_{g,\mathrm{cell}}/M_{min}$ that a part of its gas mass is transformed into a star particle of $M_{\mathrm{min}}$ that represents a star cluster. Here, $\epsilon_{\mathrm{SF}}=0.01$ is the star formation efficiency, $m_{g,\mathrm{cell}}$ is the mass of gas in the considered cell, and $M_{\mathrm{min}}=1000$ $M_\odot$ is the minimum mass of a star particle (Goldbaum et al., 2014, in preparation). \red{This simulation did not include supernova feedback, both in order to limit the computation time and to provide a baseline for the effects of gravitational instability alone, without the extra turbulence provided by supernovae. Simulations including supernovae are in progress, and will be described in future work.}
	    
	   Our initial conditions follow the setup for isolated galaxies defined by the AGORA project \citep{Kim2014}. Full details on the simulation setup are given in \citet{Kim2014}, \citet{Springel2005}, and Goldbaum et al.~(2015, in preparation), so we simply summarize here the most relevant properties. We start with a dark matter halo taken from the AGORA project, with a circular velocity of $240$ km s$^{-1}$. We initialize the baryons as a cylindrical gas cloud with a mass $M_g =4.3 \times  10^{10} M_{\odot}$ in a gaseous halo with the same mass. The gas fraction of the galaxy is 20\%. The remaining mass is composed of star particles. The gas density in the galaxy decreases exponentially both in radius and in height, i.e., the original gas profile is
	\begin{equation}
		\rho_g(r,z) = \rho_{g,0}\exp\left(-\frac{r}{R_g}\right) \exp\left(-\frac{|z|}{h_g}\right),
	\end{equation}		
where $\rho_{g,0}=10^{-23}$ g cm$^{-3}$ is the initial gas density in the center of the galaxy and $h_g=343$ pc and $R_g=3.43$ kpc are the initial vertical and radial scale length. The gas density follows this profile until it reaches the halo. The boundary between the halo and the galaxy is determined by the condition
	\begin{equation}
		\rho_gT_g =\rho_{h,0}T_h,
	\end{equation}	
where $\rho_{h,0} =10^{-30}\mbox{g cm}^{-3}$ is the density of the halo, $T_g=10^4$ K is the initial temperature of the disc, and $T_h=10^6$ K is the initial temperature of the halo. The density and temperature within the halo are uniform. The velocity in the disc is initially purely circular and fits a flat rotation curve at large radius, giving an orbital period $t_{\mathrm{ orb}} = 175 \ \mathrm{ Myr} $ at $8\ \mathrm{kpc}$. The velocity is equal to zero in the halo initially. 
	  
	  We place the galaxy in a computational box formed by a $64^3$ cube root grid with 10 levels of refinement  by a factor of 2 per level. The cell size on the finest level is $d x_{10}=20$ pc, so we are able to marginally resolve the scale height of the galaxy after the gravitational collapse. The grids are refined on criteria of gas mass in a cell, particle mass in a cell and Jeans length \citep{Truelove97}. We resolve the Jeans length by at least 32 cells on all levels except the finest level; on the finest level we add artificial pressure support to ensure that the Jeans length is always resolved by at least 4 cells. We also refine at the beginning to ensure the accuracy of the initial conditions (Goldbaum et al., 2014, in preparation). 
	   
\begin{figure*}
		\begin{center}
			\includegraphics[width=0.8 \textwidth]{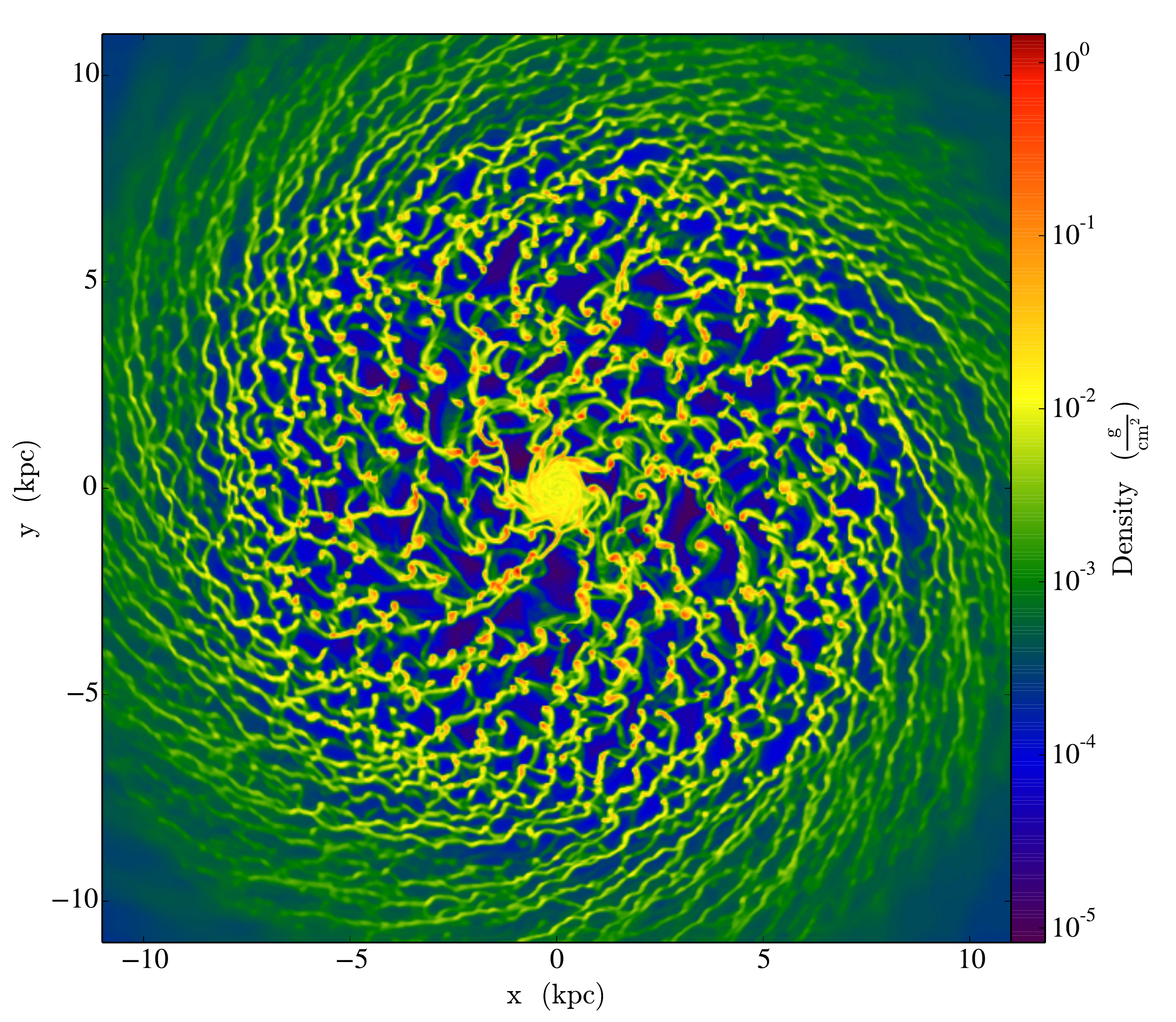}
			\caption{Projection of the density of the inner turbulent region of the galaxy.}
			\label{density}
		\end{center}
		\end{figure*}

	   To create a turbulent thin disc, we first run the simulation for 150 Myr. Over this timescale the gas collapses into a thin disc with a height of 200 pc, and a turbulent region in the center with a roughly 10 kpc radius appears. See Goldbaum et al.~(2014, in preparation) for full details on the evolution of the galaxy during this time. The density distribution of the inner region of the galaxy after this 150 Myr of evolution is displayed in Figure \ref{density}. 

	\subsection{The metal tracers}

		To study how the turbulence mixes metals in a galaxy, we take the simulation as it stands after 150 Myr of evolution and add a set of passive scalar fields, or ``colors''. The initial conditions for a realistic metal field depend
on the full star formation and merger history of a galaxy, and are obviously not available to us for our artificial initial conditions. Even in a fully cosmological simulation there might be very significant variations in metal distribution from one galaxy to another. However, we can overcome this problem by initializing the tracer with a very general pattern. If we consider a field $\chi(r,\theta)$ defined on a disc that goes to zero (or any other constant, since the offset does not change the mixing) at the edge, $\chi$ can be decomposed in a sum of the product of Bessel and trigonometric functions
		\begin{equation}
			\chi(r,\theta)  =  \sum_{n=1}^\infty \sum_{m=0}^\infty  \left[ a_{nm} J^c_{nm}(r,\theta)	 + b_{nm} J^s_{nm}(r,\theta)\right],
			\label{FBdecomposition}
		\end{equation}				
		where $J_{nm}^c$ and $J_{nm}^s$ are the Fourier-Bessel functions defined by
		\begin{gather}
			J^c_{nm}(r,\theta) = J_m(z_{nm}r/R)\cos(m\theta) \\
			J^s_{nm}(r,\theta) = J_m(z_{nm}r/R)\sin(m\theta),
		\end{gather}
		$a_{n,m}$ and $b_{n,m}$ are coefficients defined by
		\begin{gather}
			a_{nm} = \frac{\langle\chi|J^c_{nm}(r,\theta)\rangle}{||J^c_{nm}(r,\theta)||^2}	\\
			b_{nm} = \frac{\langle\chi|J^s_{nm}(r,\theta)\rangle}{||J^s_{nm}(r,\theta)||^2},
		\end{gather}				
		$R$ is the radius of the disc we are considering, $z_{nm}$ is the $n$th positive zero of $J_m$, and $J_m$ is the $m$th Bessel function of the first kind. The inner products appearing in the equation above are defined by
		\begin{equation}
			\langle f|g \rangle = (\pi R^2)^{-1}\int f g \,dA = \int_0^{2\pi}\int_0^R \frac{f g}{\pi R^2} r \,dr \,d \theta.		
			\label{IP}
		\end{equation}
		Note that the functions are $J^c_{nm}$ and $J^s_{nm}$ are orthogonal under this inner product, i.e., $\langle J^c_{nm} | J^c_{n'm'}\rangle \propto \delta_{nn'} \delta_{mm'}$ and similarly for $J^s_{nm}$. Furthermore, recall that $J_m(z_{nm}r/R,\theta)\cos(m\theta)$ and $J_m(z_{nm}r/R,\theta)\sin(m\theta)$ are eigenvalues for the Laplacian. Thus we would expect that, for a pure diffusion process, the evolution of any initial metal field $\chi$ could be described simply as a decrease in amplitude of the various Fourier-Bessel modes into which it can be decomposed, with no transfer of power between them. In this respect using a Fourier-Bessel decomposition is the natural extension to cylindrical coordinates of the decomposition of a metal field into Fourier modes for a shearing periodic box introduced by \citet{Yang2012}.
		
Given this analysis, to study mixing we take the state of the simulation at 150 Myr and add 25 different color fields with the following initial conditions:
		\begin{equation}
		 	\chi_{nm}(r,\theta,z,t=0) = J_m(z_{nm} r/R)\cos(m\theta)= J^c_{nm}(r,\theta),
		 	\label{chinm}
		\end{equation}
where $R = 8 $ kpc is the radius of the turbulent area, $n=1\ldots 5$ and $m=0\ldots 4$, and $(r,\theta,z) \in [0,R] \times [0,2\pi] \times [-h_g,h_g]$; $\chi_{nm}$ is set to zero outside of this cylinder. We only consider the cosine terms because the fields represented by the sine terms are identical up to a 90$^\circ$ rotation about the $z$ axis. Since the problem is, in a statistical sense, invariant under such a transformation, the modes represented by the sine terms should not mix any differently than the ones represented by the cosine terms, and thus there is no additional information gained by including them.

\red{Once they are added, the color fields are advected by the gas motion as passive scalars. Numerically, we handle advection of metal tracers in exactly the same manner as advection of mass. Thus if the hydrodynamic solver returns a mass flux $F_M$ across a particular cell face, the flux of ``metal mass" in field $n,m$ is simply $\chi_{nm,\mathrm{upwind}} F_M$, where $\chi_{nm,\mathrm{upwind}}$ is the upwind value of the metal field $\chi_{nm}$, determined using the same PPM interpolation scheme as for the hydrodynamic quantities. This scheme is conservative, in the sense that hydrodynamic evolution leaves the total ``metal mass" in the simulation domain, $\int \rho \chi_{nm}\,dV$, unchanged, even as the color field is advected between computational cells. When stars form, we leave the color concentration $\chi_{nm}$ in the star-forming cell unchanged, which amounts to assuming that the metal concentration in the stars is identical to that of the gas cells in which they form. In principle we could then track the metal contents of the stars \citep[e.g.,][]{feng14a}, but we have not done so here, because our simulation does not run long enough for significant stellar migration to occur, and it is not clear if our resolution of the $N$-body dynamics of the stars is sufficient to follow stellar migration accurately in any event.
}

\red{
We pause to make two final points regarding our numerical method. Some previous simulations of metal transport in isolated galaxies have used smoothed particle hydrodynamics (SPH) methods \citep[e.g.][]{kubryk13a, grand15a}. These codes have an advantage over our method in that, because they allow the authors to track individual particles, it is possible to distinguish between ``radial migration'' of gas, i.e., bulk radial transport of gas, and ``mixing'', i.e., homogenization of the metal distribution but without bulk radial redistribution of mass. We cannot make this distinction, and thus we will use the generic term mixing to refer to any process that homogenizes the metal distribution.}

\red{On the other hand, our Eulerian method also offers a significant advantage over SPH when it comes to simulation of metal transport. SPH codes require a prescription for sub-grid scale turbulent mixing in order to follow metal transport, because without such a model SPH artificially suppresses mixing below the resolution scale \cite[e.g.,][]{wadsley08a, greif09a}. These sub-grid models contain a number of free parameters, and their accuracy has not been extensively calibrated. In contrast, our simulations mix naturally at the grid scale, and do not require us to use an explicit subgrid mixing model.
}

\section{Analysis: axisymmetric and non-axisymmetric modes}
\label{analysis}

	After inserting the color fields as described above, we allow the simulation to evolve for another \mixtime. During this time each of the passive scalar fields $\chi_{nm}$ evolves, and in this section we discuss the nature of this time evolution. We have available the value of each scalar field as a function of time and position, $\chi_{nm}(r,\theta,z,t)$, but in order to simplify the analysis we neglect their vertical structure and only analyze their column density-weighted averages. Specifically, we define
		\begin{equation}
			\chi_{nm}(r,\theta, t) = \frac{ \int\chi_{nm}(r,\theta,z,t)\rho(r,\theta,z,t) \, d z}{\int\rho(r,\theta,z,t)\, d z},
		\end{equation}
	where $\rho$ is the gas volume density, and from now on when we refer to $\chi_{nm}$ we mean the vertically-integrated quantity. We perform the integration, and all other analysis presented in this paper, using the analysis tool \texttt{yt} \citep{Turk2011}. \red{Since our simulations do not include supernova feedback, the disk remains thin and there is no gas expulsion on the vertical direction. For this reason, transport of tracers along the vertical direction is negligible and the integration over $z$ does not significantly alter the results.}

	We quantify the evolution of the color fields by computing the projection of $\chi_{nm}(r,\theta,t)$ on the different Fourier-Bessel basis functions. \red{The projection is done on a disc of radius $R = 8 \ \mathrm{kpc}$ with the inner product defined on equation \ref{IP}}. We define these projections by
					 \begin{eqnarray}
		 	P^c_{nm,n'm'}(t)& = &\langle\chi_{nm}(t)|J^c_{n'm'}\rangle \\
		 	P^s_{nm,n'm'}(t)&=&\langle\chi_{nm}(t)|J^s_{n'm'}\rangle.
		 \end{eqnarray}
		The quantity $P^{s,c}_{nm,n'm'}(t)$ indicates how much of the initial power from the mode $n,m$ has been transferred to the mode $n',m'$ (or, for $n=n'$ and $m=m'$, how much of the original power remains) at time $t$. Note that, even if the metal field remained in a fixed pattern and the galaxy rotated as a solid body, this rotation would exchange power between the $J_{nm}^c$ and $J_{nm}^s$. To remove this effect, rather than considering $P^c_{nm,n',m'}$ and $P^s_{nm,n'm'}$ separately, we instead compute $\P_{nm,n'm'}$, where
		\begin{equation}
			\P^2_{nm,n'm'} =\left(P^c_{nm,n'm'}\right)^2+\left(P^s_{nm,n'm'}\right)^2.
			\label{Pnm}
		\end{equation}			
	This quantity is invariant under rotation, so if the galaxy rotated as a solid body and the spread of the metal field were described simply by a diffusion equation in the rotating frame, then we would have $\P_{nm,n'm'}(t) = \delta_{nn'} \delta_{mm'} f(t)$, with $f(0)=1$ and $f(t)$ strictly decreasing with $t$ for $t>0$. In reality we shall see that this is not how the color fields evolve, but this case nonetheless provides a useful zeroth order model against which our findings can be compared.

	\subsection{Axisymmetric metallicity variation}
		
		A realistic metallicity field will contain contributions from
 two different types of modes: the axisymmetric ones, i.e., those with $m=0$, carry information on the average radial dependence, and the non-axisymmetric ones, i.e., the $m\neq0$ modes, which describe spiral structure or other deviations from uniformity at fixed radius. Since 2-armed spirals are the most common type of non-axisymmetric pattern in galaxies, we will focus on the case $m=2$. We discuss the $m=0$ modes in this Section, and the $m=2$ ones in the subsequent one.

		\begin{figure*}
		\begin{center}
			\includegraphics[width=0.8 \textwidth]{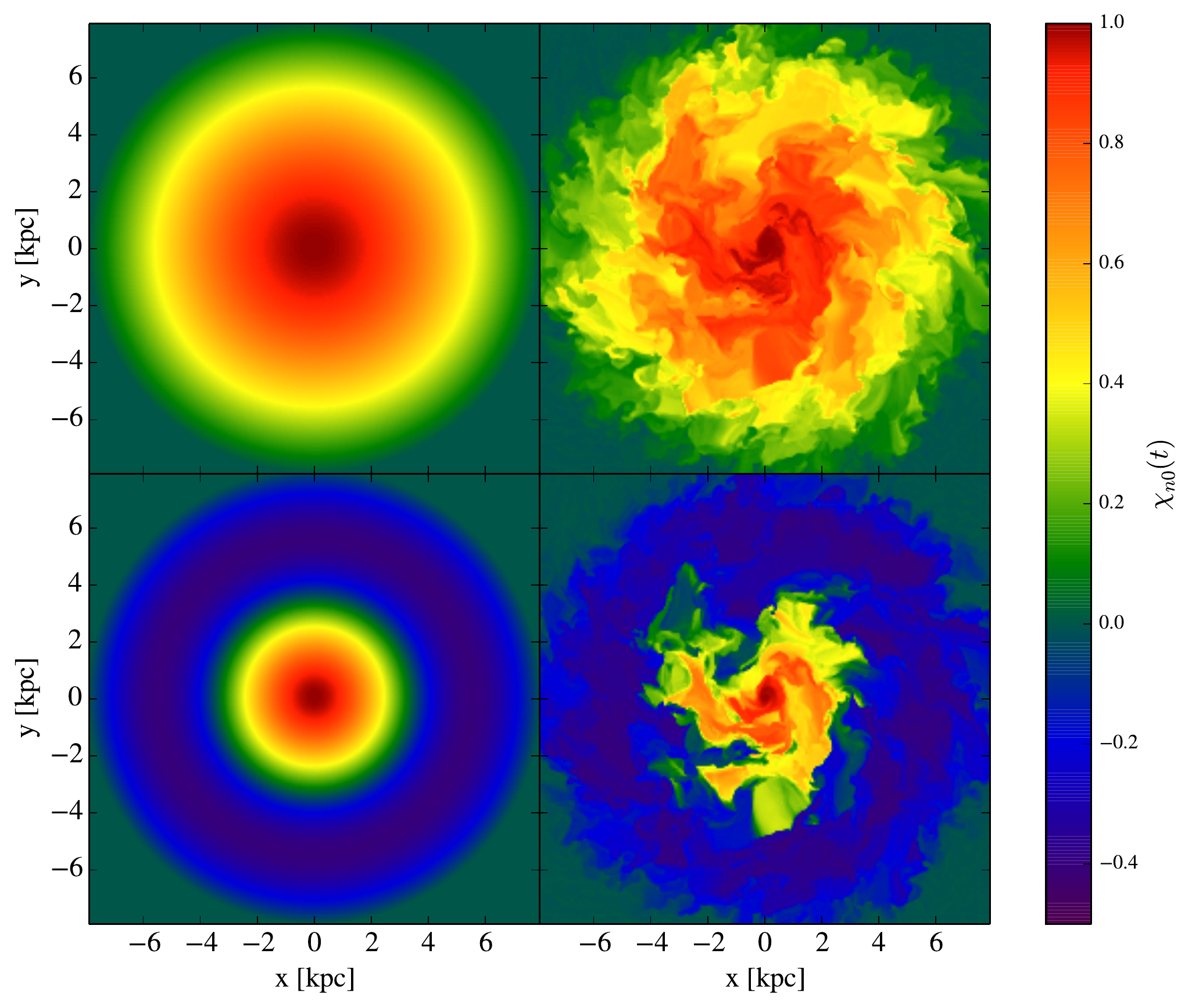}
			\caption{Snapshots at $t_0$ (left) and $t=t_0+\mixtimem$ Myr (right) of the $m=0,$  $n=1$ (top) and $m=0$, $n=2$ (bottom) color fields $\chi_{nm}$.} 
			\label{m=0snap}
		\end{center}
		\end{figure*}

		Figure \ref{m=0snap} shows the vertically-integrated color fields $\chi_{nm}$ for $m=0$, $n=1,2$ at $t=t_0$ (the time when the color fields are first added) and $t=t_0+\mixtimem$. As a consequence of the choice of $R$, the gas in which the tracers are deposited is fully turbulent. The original pattern is slowly destroyed by the turbulence, and the destruction looks faster for $n=2$ than for $n=1$.

\begin{figure}
			\begin{center}
				\includegraphics[width=0.5 \textwidth]{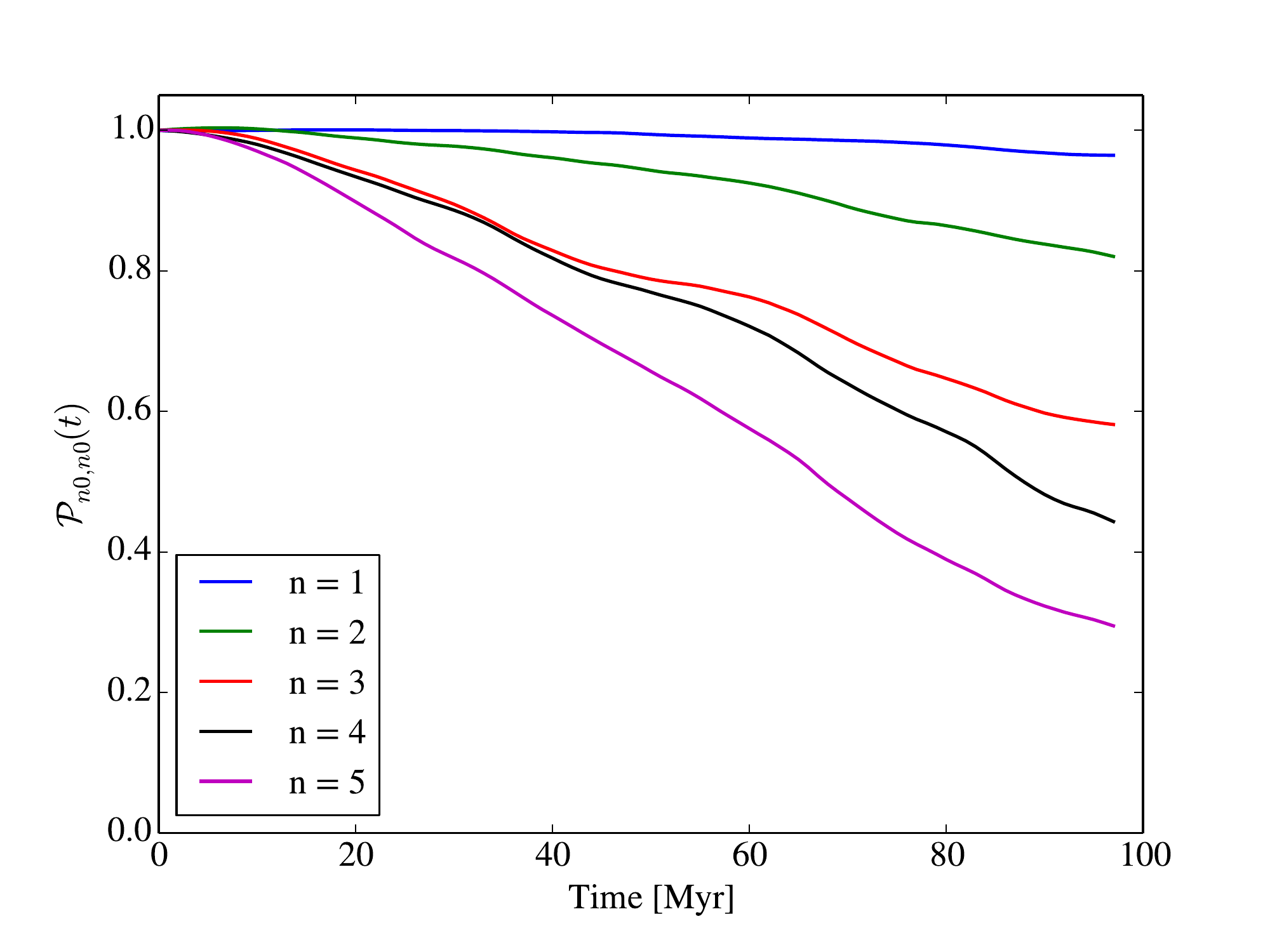}
				\caption{Evolution over time of $\P_{n0,n0}(t)$, the fraction of the original power remaining in each of the $m=0$ modes (equation \ref{Pnm}) for $n$ varying from 1 to 5.}
				\label{Pn0n0}
			\end{center}
		\end{figure}

		Confirming this visual impression, Figure \ref{Pn0n0} shows the decrease of $\P_{nm,nm}(t)$, the power remaining in the original mode, with time for the axisymmetric modes $n \in \{1 \ldots 5\}$. We observe that the mode $m=0$, $n=1$ remains almost completely stationary for the time scale we consider, while the other modes slightly decrease. The rate at which the modes decrease is inversely correlated with $n$, which is not surprising: higher $n$ modes correspond to smaller spatial scales, and we expect that turbulence should mix out smaller-scale inhomogeneities faster than larger-scale ones. After \mixtime , the mode that has decreased the most, $n=5$  has lost roughly 70\% of its original power, but it is still in a linear phase.
			
\begin{figure}
			\begin{center}
				\includegraphics[width= 0.5 \textwidth]{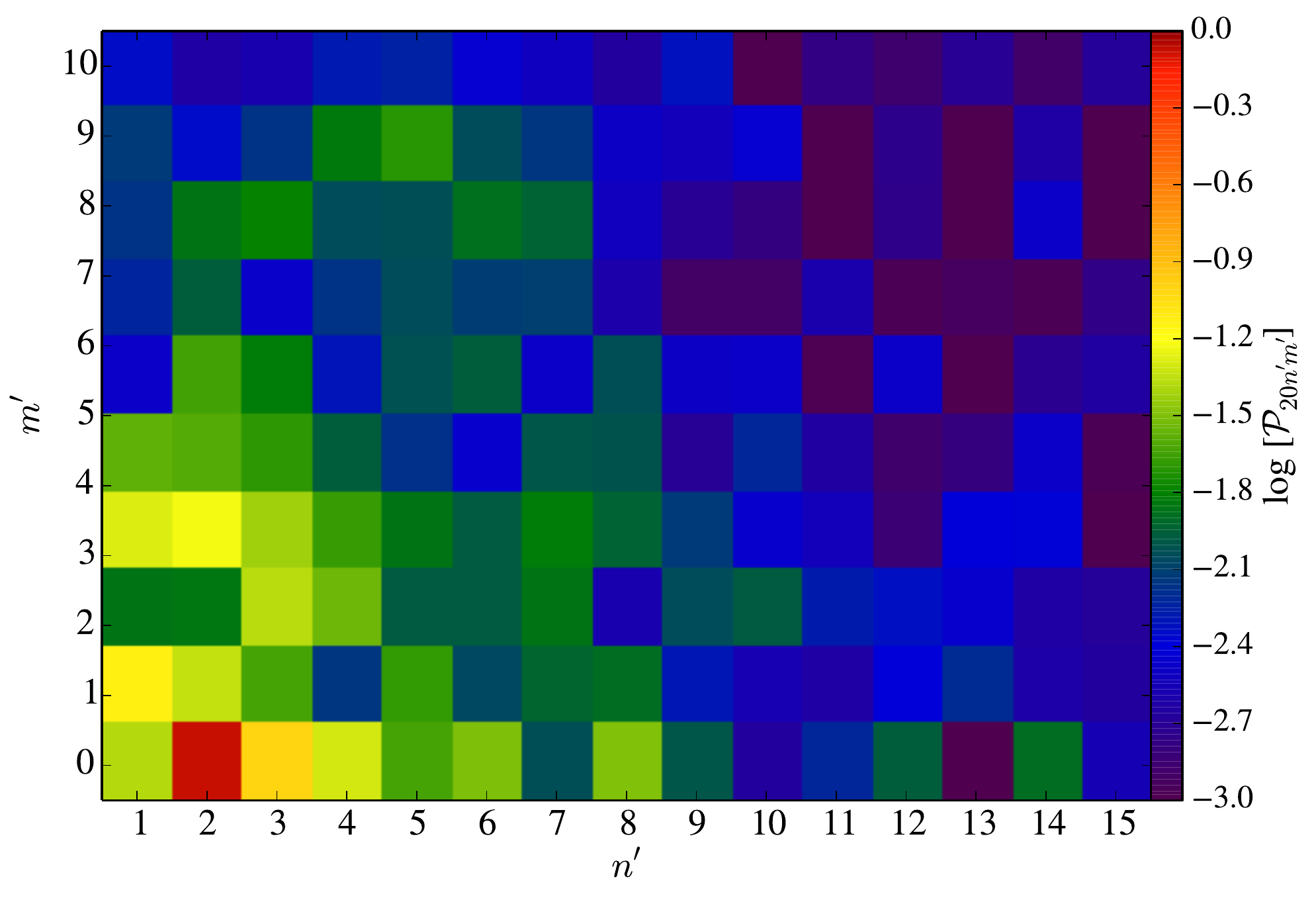}
				\caption{Power transferred from $n=2, m=0$ to all modes, $\P_{20,n'm'}$, for $n'$ varying from 1 to 15 and $m'$ from 0 to 10 after \mixtime\ of mixing}
				\label{FBm=0n=2}
			\end{center}
			\end{figure}

		We can next investigate how power is transferred between modes, which is described by the quantity $\P_{nm,n'm'}$(t). In Figure \ref{FBm=0n=2}, we show $\P_{20,n'm'}(t)$ at $t=t_0+\mixtimem$. Physically, this quantity shows how the power that was originally in the $n=2$, $m=0$ mode has been transferred to other modes over \mixtime\ of evolution. We can see from the figure that turbulent diffusion transfers power to both higher $n$ and higher $m$, and that it does so approximately isotropically. This figure is also consistent with the result shown in Figure \ref{Pn0n0} that the mixing is  slow for the $n=2$, $m=0$ mode. The majority of the power remains in the original mode, and only $\sim 10\%$ percent of it has spread to other modes.  \red{Recall, however, that, at the time illustrated in Figure \ref{FBm=0n=2}, the simulation has evolved for only \mixtime, which is less than a full orbital period of the disc (at 8 kpc). It is therefore not surprising that only a relatively small fraction of the total power has been transferred out of the mode}.

	\subsection{Non-axisymmetric metallicity variation}
	
\begin{figure*}
		\begin{center}
			\includegraphics[width=0.8 \textwidth]{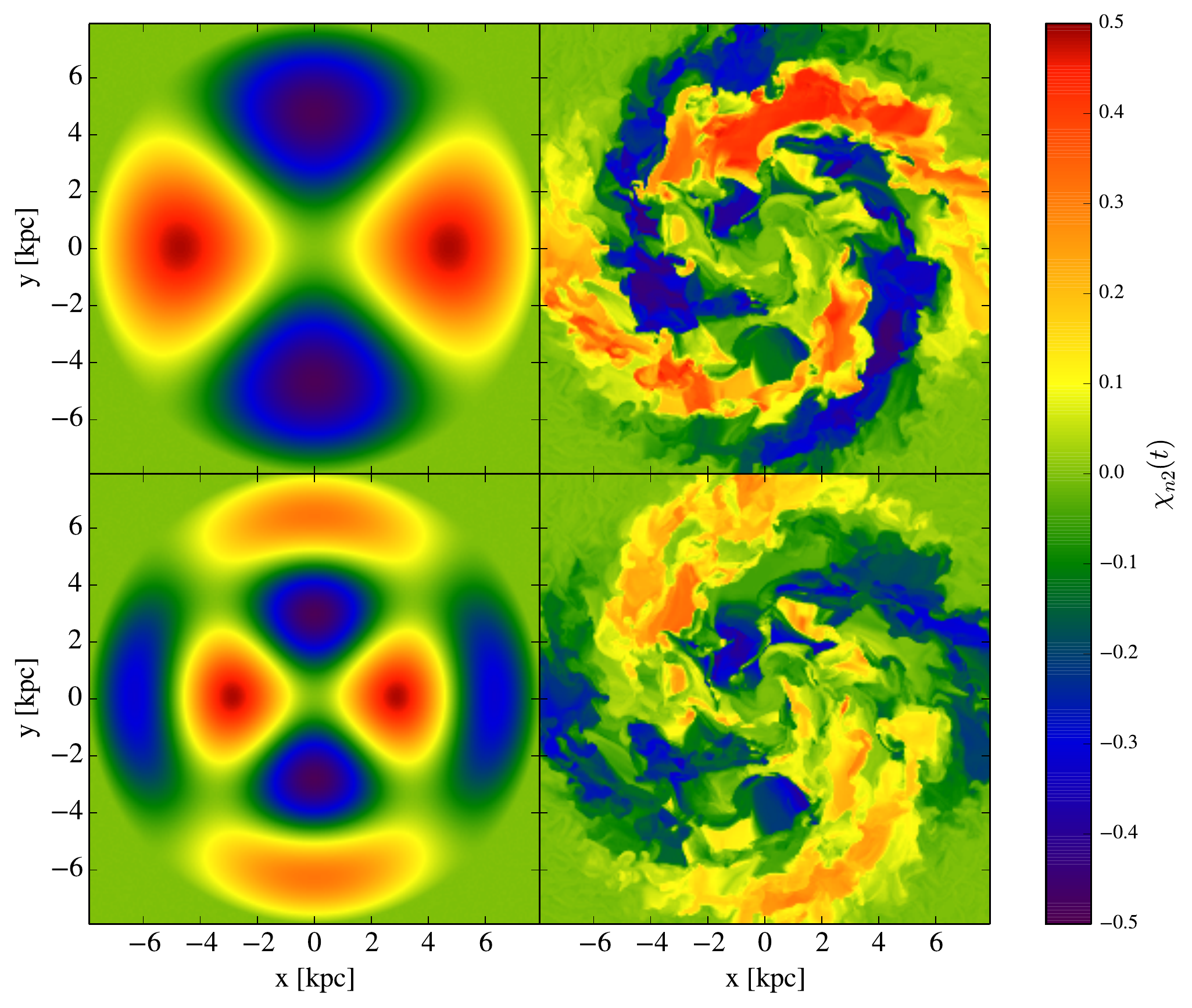}
			\caption{
			Same as Figure \ref{m=0snap}, but now for the $m=2$, $n=1$ (top) and $m=2$, $n=2$ (bottom) color fields.
			\label{m=2snap}
			}
		\end{center}
		\end{figure*}

		Star formation that is concentrated in spiral arms is likely to produce non-axisymmetry in the metal field, and we therefore next consider non-axisymmetric modes. Which mode dominates in a real galaxy will likely depend on whether the spiral arms are grand design or flocculent. \red{For this} example we choose to focus on $m=2$, the mode that should dominate for a two-armed spiral. We present results for all other $m$ modes below.  In Figure \ref{m=2snap}, we show some snapshots of the modes $m=2$, $n=1,2$ at $t_0$ and $t_0+ \mixtimem$. We can see that the mixing looks more efficient than in the axisymmetric case. The  pattern has been destroyed both by the turbulence and the differential rotation. We will study in this part both of these mechanisms.
		 
\begin{figure}
			\begin{center}
 				\includegraphics[width= 0.5 \textwidth]{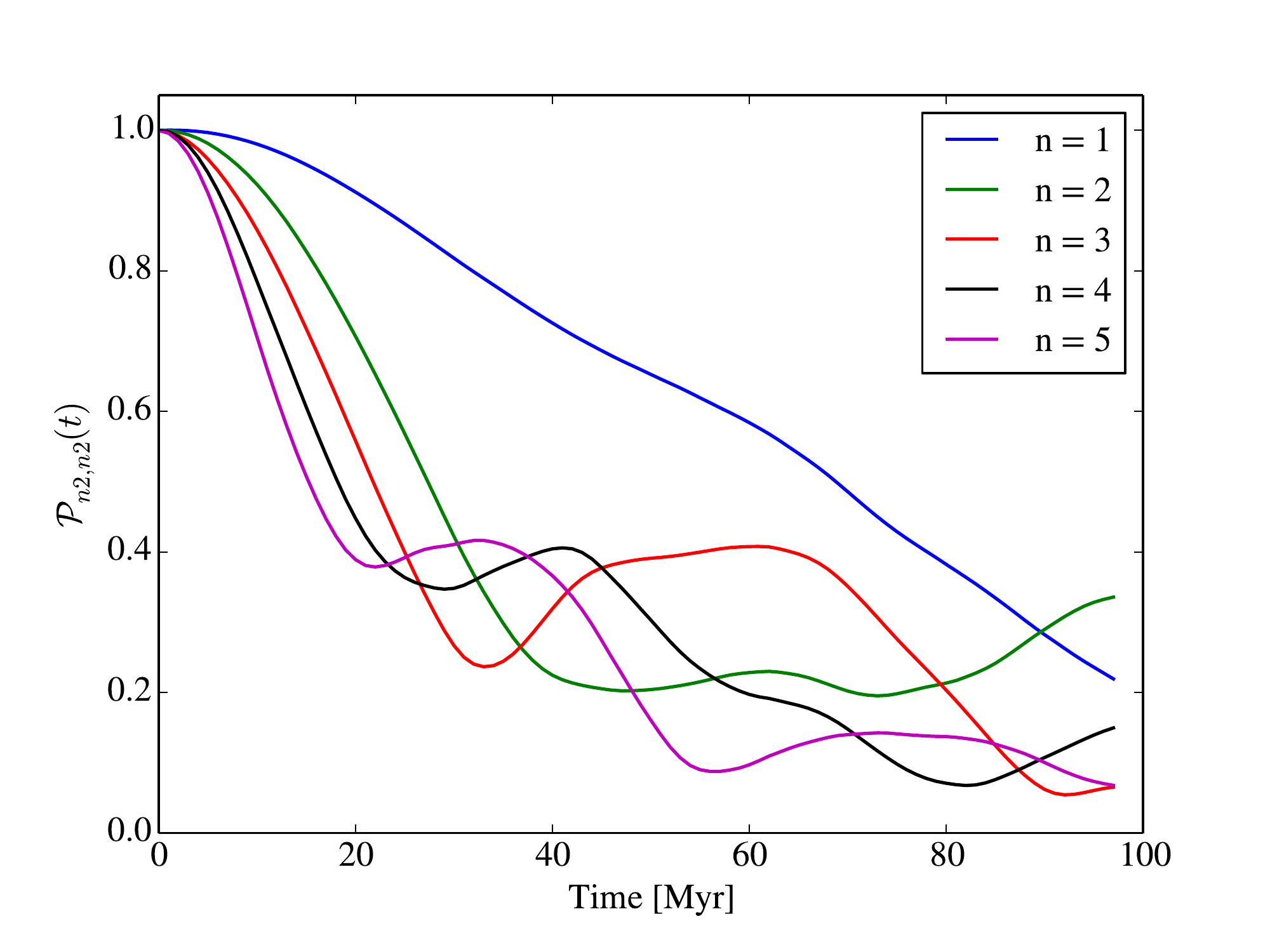}
				\caption{
				Same as Figure \ref{Pn0n0}, but for the $m=2$ rather than $m=0$ modes.
				\label{Pn2n2}
				}
			\end{center}
			\end{figure}

\begin{figure}
				\includegraphics[width= 0.5 \textwidth]{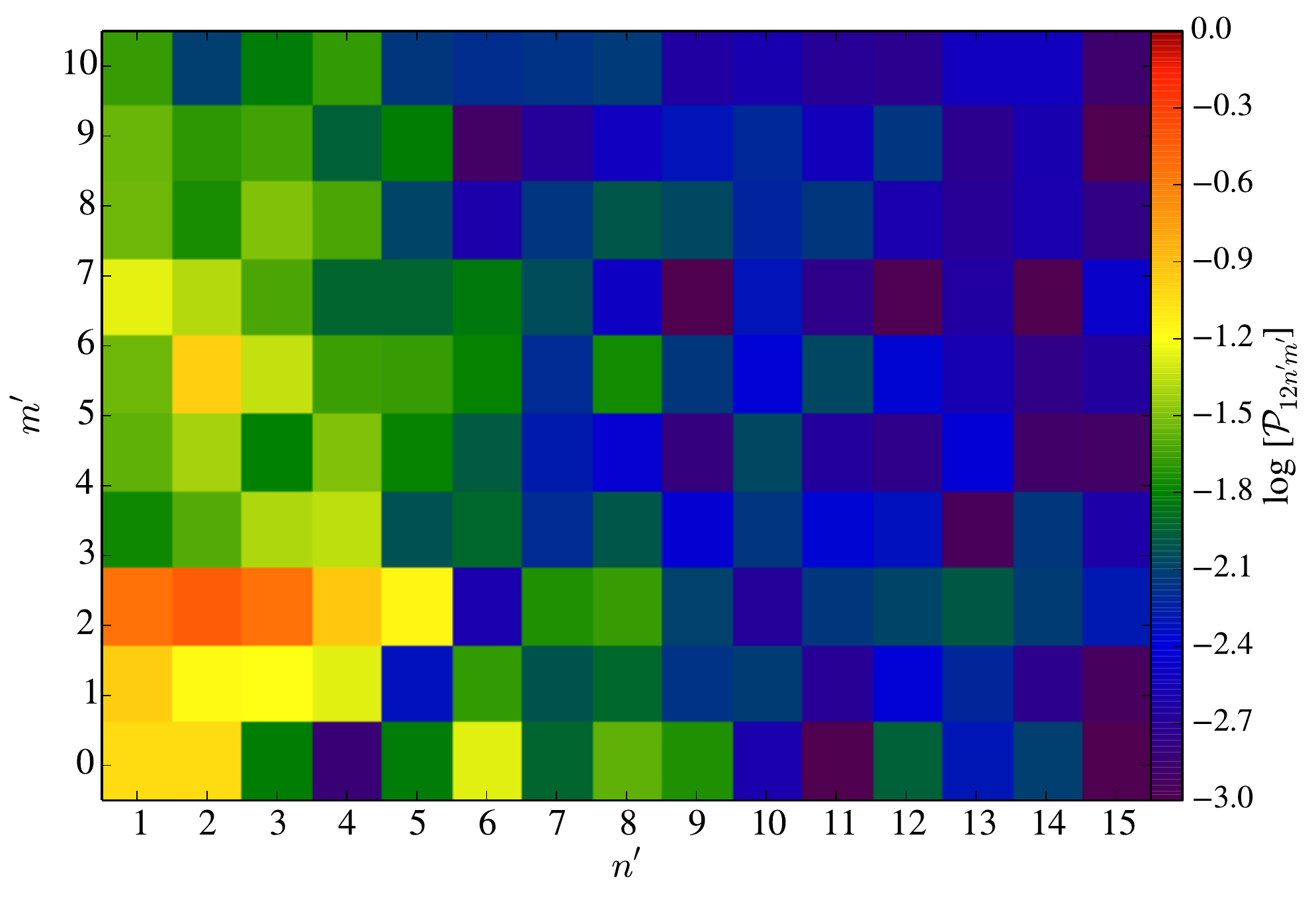}
				\caption{
				Same as Figure \ref{FBm=0n=2}, but for the $n=1$, $m=2$ modes.
				\label{FBm=2n=1}
				}
			\end{figure}

		In Figure \ref{Pn2n2}, we can see that, after a brief transient, $\P_{nm,nm}(t)$ decreases linearly with time before oscillating between 0 and 0.2 of the original value for the higher values of $n$. The modes vanish much faster than the axisymmetric ones.  However, most of the reduction in power is a result of transfer to other modes rather than outright destruction. Indeed, if we compute the sum of the power remaining over all the modes, $\sum_{n',m'} \P^2_{nm,n'm'}(t)$, which is equivalent to computing the norm in the real space $||\chi_{nm}(t)||^2$, we find that it decreases by at most of order $20\%$ over the \mixtime\ of evolution.\footnote{We can estimate the maximum values of $n$ and $m$ accessible to our simulation from our resolution of 20 pc. Since the disc radius is 8 kpc, we have roughly 400 cells per disc radius. If we assume that we need $\sim 10$ cells per wavelength to resolve the mode, then we should be able to obtain reliable estimates up to $n$ and $m$ values of $\sim 40$.} Figure \ref{FBm=2n=1}, which shows $\P_{12,n'm'}$ evaluated at $t=t_0+\mixtimem$, reveals where the power that is removed from the $n=1$, $m=2$ mode is transferred. We can see that most of the power is still in modes with $m=2$, while a smaller fraction has leaked into $m\neq2$ modes (about 20\%). 
		
		\subsubsection{The shear}

\begin{figure}
				\begin{center}
					\includegraphics[width= 0.5 \textwidth]{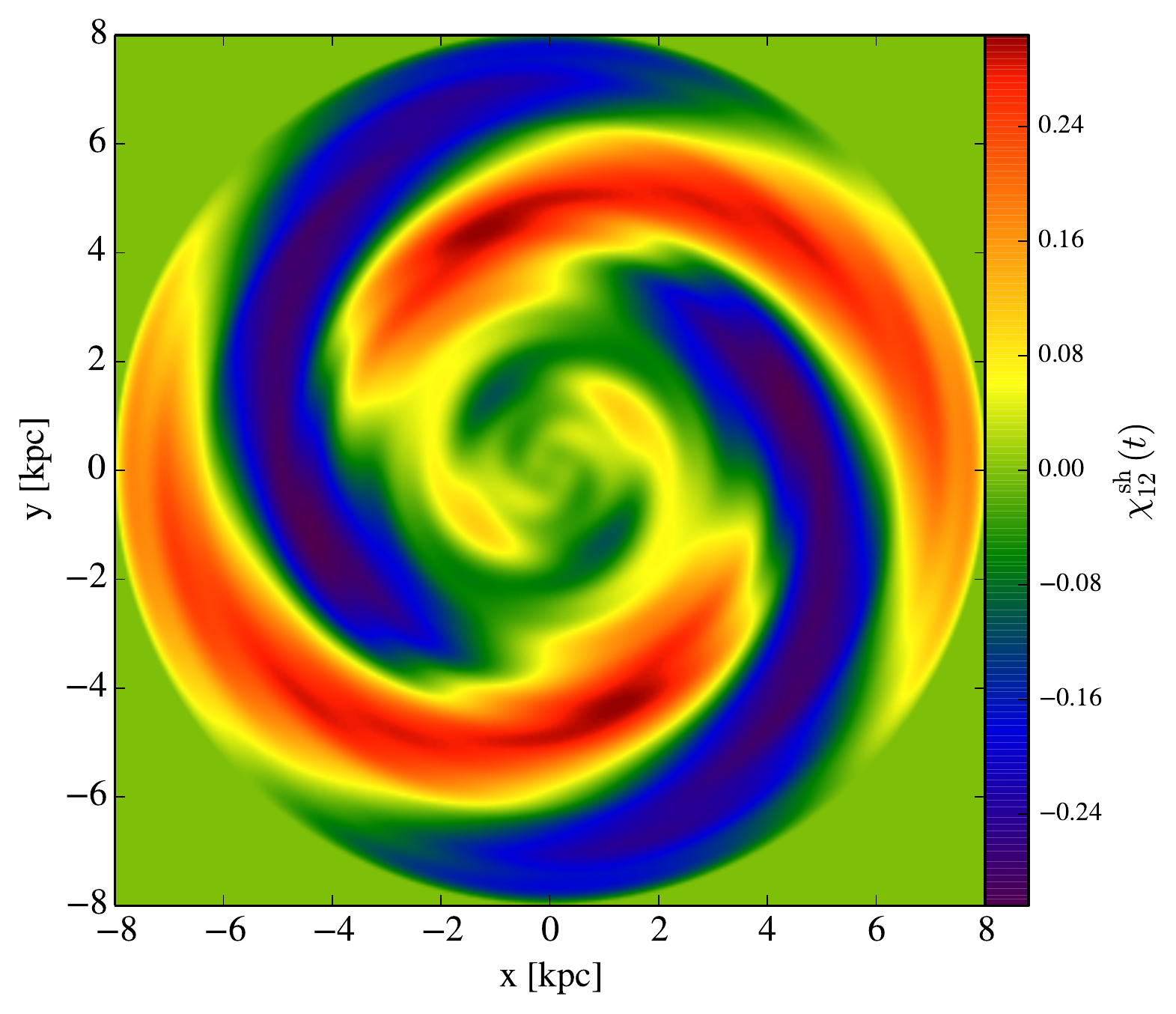}
					\caption{Projection of $\chi_{12}(t)$ after \mixtime\ of mixing on the modes $m=2$, $n=1..15$ using $P^{c,s}_{12,n'm'}$ \red{as defined in equation \ref{sheareq}}. It reproduces the shear induced by the \red{differential} rotation of the galaxy.}
					\label{shear}
				\end{center}
			\end{figure}

			The main cause of the spread of the power in the $m=2$ line shown in Figure \ref{FBm=2n=1} is the shear induced by the differential rotation. We illustrate this point in Figure \ref{shear}, which shows the real-space reconstruction of the color field that results from summing up only the $m=2$ modes in Figure \ref{FBm=2n=1}, i.e.,
				\begin{equation}
				\chi^{\textrm{sh}}_{n 2}(t)=\sum_{n'}\left[{\frac{P^c_{n2,n'2}(t)}{||J^c_{n'2}||^2}J^c_{n'2}+\frac{P^s_{n2,n'2}(t)}{||J^s_{n'2}||^2}J^s_{n'2}}\right].
				\label{sheareq}
			\end{equation}						
	As is clear from Figure \ref{shear}, this procedure picks out the sheared pattern created by the differential rotation. As first pointed out by \citet{Yang2012}, this a very important effect for diffusion. As we have already seen when considering axisymmetric modes, turbulence is much more efficient at mixing away inhomogeneities on small physical scales (higher $n$ and $m$) than on large ones. Since the primary effect of differential rotation is to transport power to higher $m$, even if the shear by itself doesn't mix efficiently, it increases significantly the speed of the diffusion because the radial pattern is not preserved by differential rotation.
		
		\subsubsection{The turbulent mixing}
		 	
			\begin{figure}
				\begin{center}
					\includegraphics[width=0.5 \textwidth]{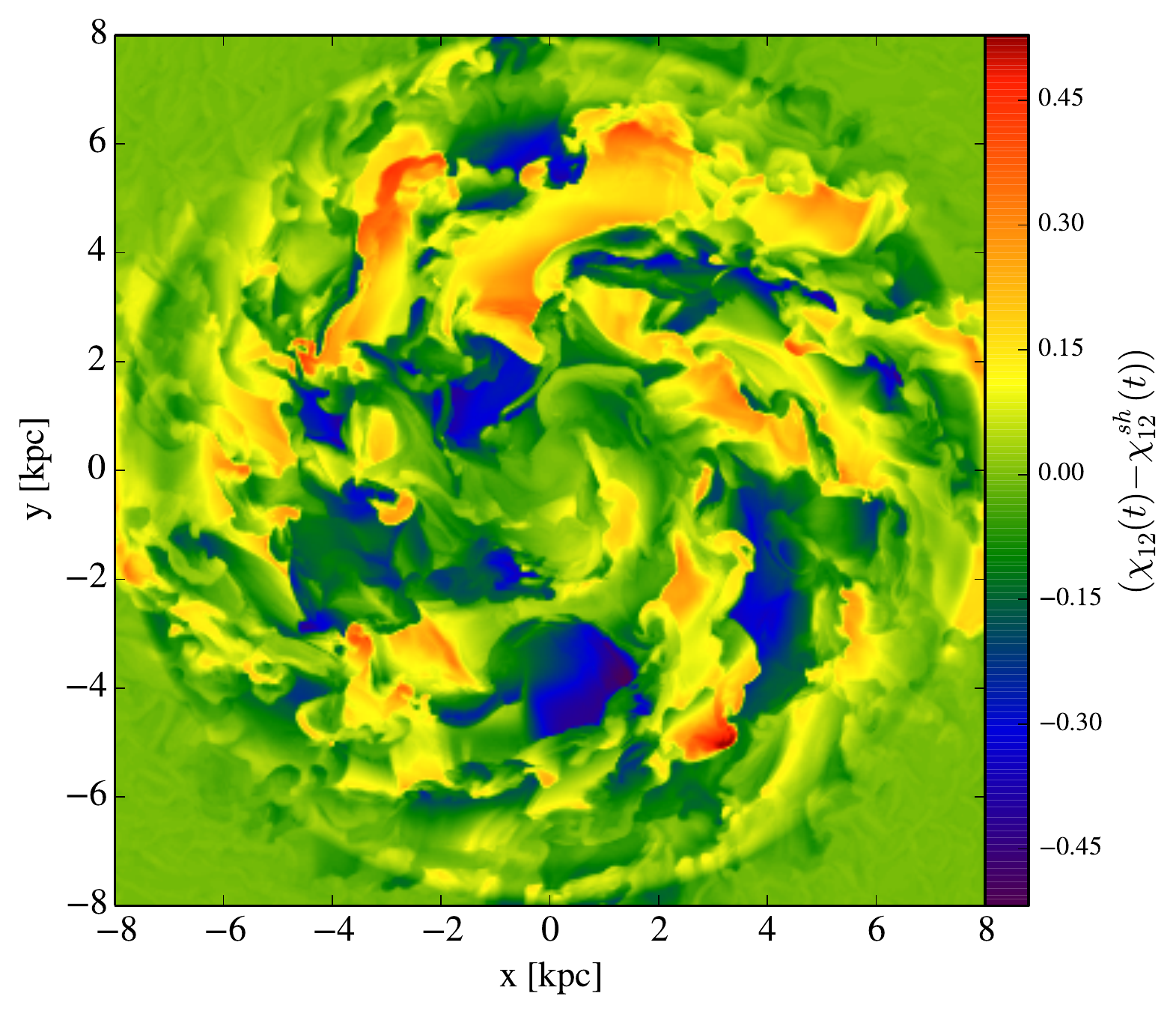}
					\caption{\red{Difference between the full color field $\chi_{12}(t=\mathrm{t_0+\mixtime})$ at time $t_0+\mixtime$, as shown in the top right panel of the figure \ref{m=2snap}, and the partial reconstruction $\chi_{12}^{\rm sh}$ defined by equation \ref{sheareq} and shown in Figure \ref{shear}. Intuitively, this field shows the change in the color field due to turbulence rather than shear.}
					\label{deltashear}
					}
				\end{center}
			\end{figure}	
									
			While the transport of power along the $m=2$ row is caused by differential rotation, the power driven to other modes is a manifestation of the turbulence. To demonstrate this, in Figure \ref{deltashear} we show the $\chi_{12}(t) - \chi_{12}^{\rm sh}(t)$  after \mixtime. This quantity is simply the residual that results when we subtract Figure \ref{shear} from the upper right panel of Figure \ref{m=2snap}. We can use the power associated with this residual field to quantify the importance of turbulent mixing.
			
			Let us consider the quantity
			\begin{equation}
				\mathcal{T}_{nm} = 1- \sqrt{\frac{\sum \limits_{n'} \P^2_{n'm,n'm}}{\sum \limits_{(n',m')} \P^2_{nm,n'm'}}}.
				\label{ratio}
			\end{equation}
			Intuitively, the numerator in the fraction is the total power in modes with the same $m$ as the original one, and thus represents the fraction of the original power that is in the sheared field. The denominator is simply the total power summed over all modes, and the ratio is therefore the fraction of the original power that remains in a coherent, shear-distorted pattern. One minus this quantity is the fraction of the original power that has been transported or destroyed by turbulence. Thus one may intuitively think of the quantity $\mathcal{T}_{nm}$  as the fraction of the original power that has been diffused by the turbulence.\footnote{While equation (\ref{ratio}) formally involves a sum over all $n'$ and $m'$ up to infinity, in practice we must of course terminate the computation at finite values of $n'$ and $m'$. We choose to stop at $n'=25$ and $m'=10$ because the sum seems to converge with these limits, as shown in the Appendix.} 
			
\begin{figure}
				\begin{center}
					\includegraphics[width=0.5 \textwidth]{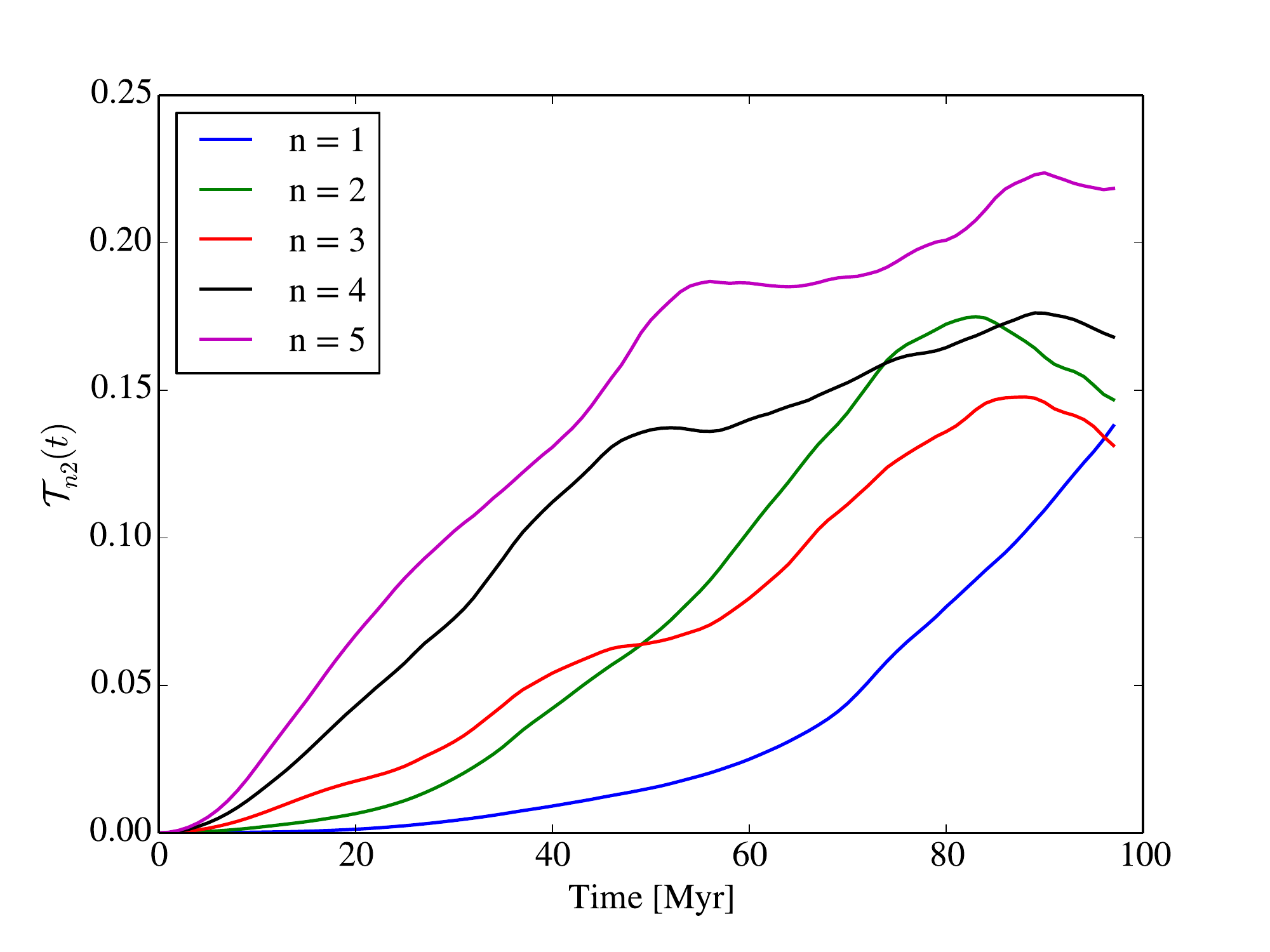}
					\caption{The power transferred from $m=2$ to $m\neq 2$ modes by turbulence, $\mathcal{T}_{nm}$  (equation \ref{ratio}), as a function of the time for $m=2$ and $n=1\ldots5$.
					\label{ratiom=2}
					}
				\end{center}
			\end{figure}

			We show $\mathcal{T}_{nm}$ for $m=2$ in Figure \ref{ratiom=2}. In the case of $m=2$, $n=5$, which has the highest dissipation rate for the modes we are considering in this part, roughly 10\% of the power is  transferred to $m\neq 2$ modes. However, this is a lower limit on the power dissipated by the turbulence, since turbulence as well as shear can transfer power between $m=2$ modes.
			
\begin{figure}
				\begin{center}
					\includegraphics[width=0.5 \textwidth]{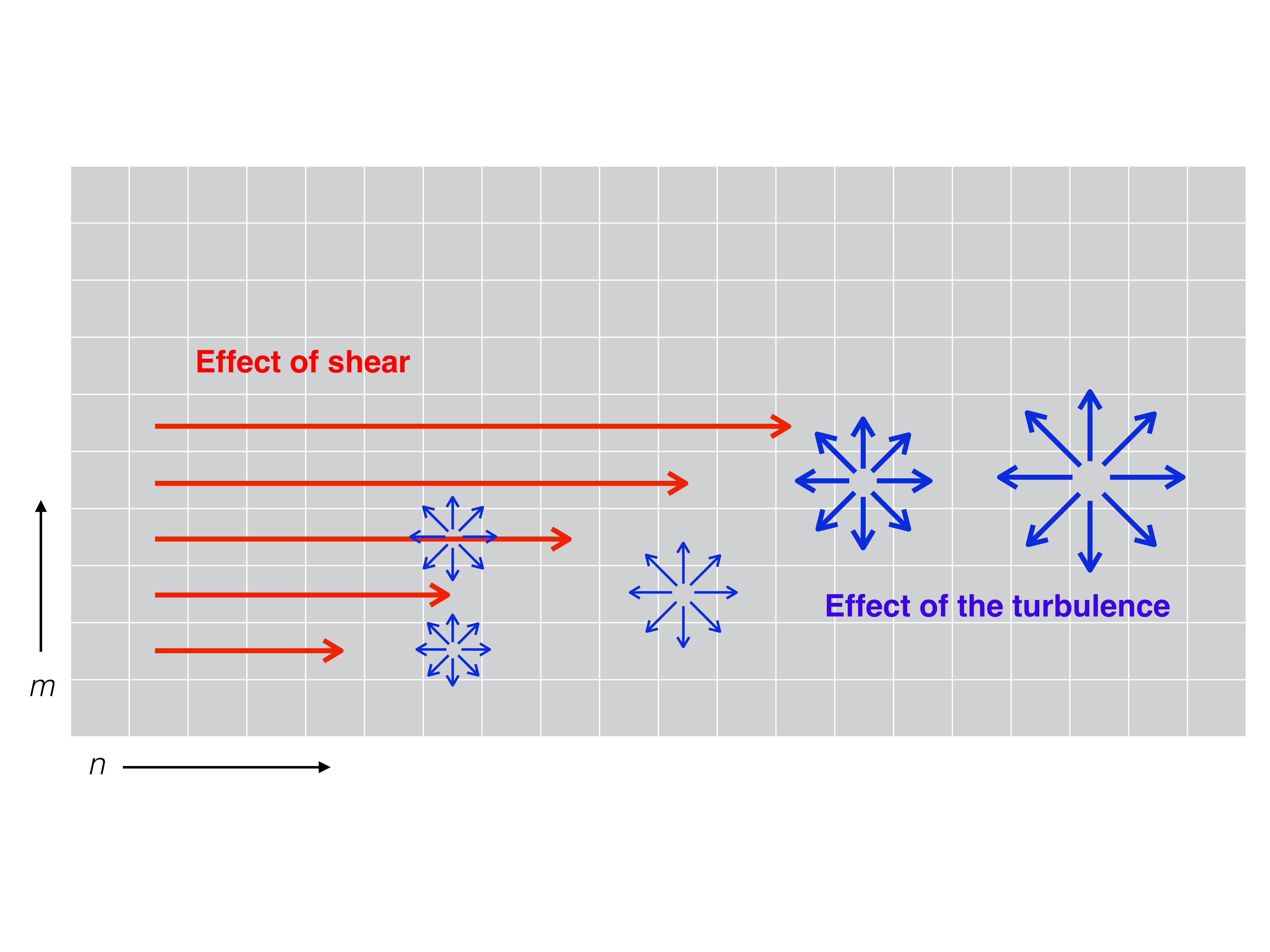}
					\caption{Schematic representation of the actions of the shear and the turbulence on the spectrum of the metal field.}
					\label{mecha}
				\end{center}
			\end{figure}

			We can schematically summarize the joint effects of turbulence and shear in Figure \ref{mecha}. Turbulence is the only agent capable of truly mixing the disc and wiping out inhomogeneities. However, it operates fairly slowly for patters with large spatial scales. In comparison, for non-axisymmetric modes, $m\neq 0$, turbulence is greatly aided by shear. Shear transports power out of low $n$ modes and into higher $n$ ones; physically, differential rotation transforms any non-axisymmetric pattern into a tightly-wound spiral on a timescale comparable to the orbital period. This works in conjunction with the turbulence to greatly increase the rate of mixing, by moving power out of low $n$ modes where it is hard to dissipate and into high $n$ modes where dissipation is more rapid \citep{Yang2012}.

	\subsection{Mixing timescales}
			
		We saw in the previous two sections that some modes can be very stable whereas others can be completely diffused on timescales well under an orbital period. We now make this analysis quantitative by associating to the destruction of each mode a timescale. Figures \ref{Pn0n0} and \ref{Pn2n2} show that the power remaining in the original mode, $\P_{nm,nm}$, is roughly constant during an initial transient, then undergoes a phase where it decays roughly linearly with time, and finally stabilizes and begins oscillating once the majority of the power is gone. This oscillation is caused by the fact that the original mode is no longer dominant and so the power from other modes can feed back into the original one. Examining comparable plots for other values of $m$, we find that these three phases are generic. It therefore seems reasonable to estimate a decay timescale for each mode $n,m$ by computing the slope $D_{mn}$ in the linear phase. $D_{nm}$ increases with $n$ and $m$ (Figures \ref{Pn0n0} and \ref{Pn2n2}), we can also notice that the slope is very small for the $m=0$ modes. The timescale is simply the inverse of this slope, i.e.,
		\begin{equation}
			\tau_{nm}=D_{nm}^{-1}.
			\label{tau}
		\end{equation}

\begin{figure}
			\begin{center}
				\includegraphics[width=0.5 \textwidth]{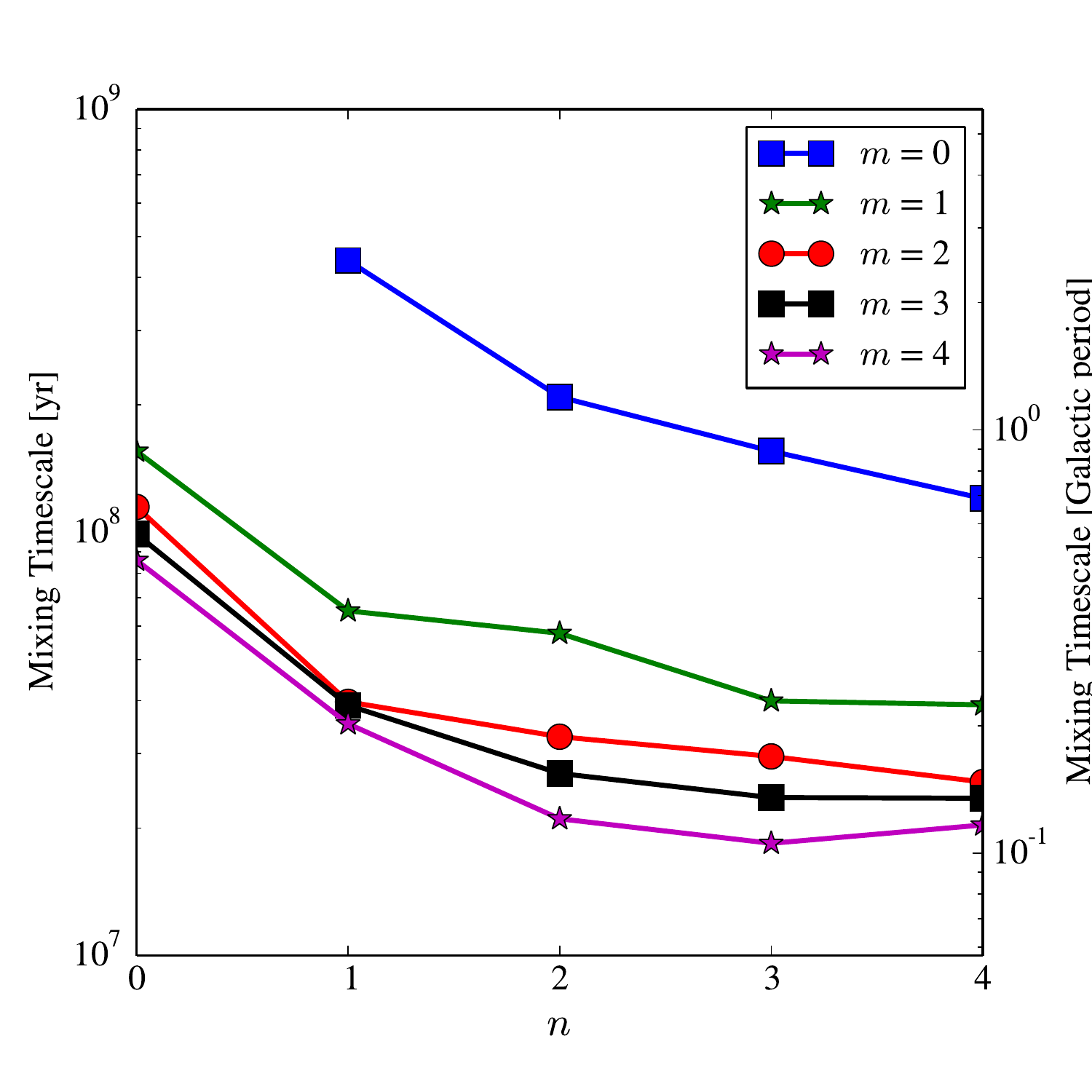}
				\caption{The timescale for metallicity inhomogeneities to decay, $\tau_{nm}$ (equation\ref{tau}), as a function of $n$ for $m=0 \ldots 4$. On the right side, we normalize the timescale to the galactic orbital period, $t_{\rm orb} = 175$ Myr at $8 \ \mathrm{kpc}$. $\tau_{10}$ is not plotted because the simulation time was too small to reach the linear phase for this mode. 
				\label{Dtimescale}
				}
			\end{center}
		\end{figure}

		Figure \ref{Dtimescale} shows $\tau_{nm}$ versus $n$ and $m$. We see that for the $m=0$ modes these times are bigger than or similar to the galactic orbital period $t_{\rm orb}=175$ Myr at $8\ \mathrm{kpc}$, whereas for the non-axisymmetric modes, all are smaller than the galactic orbital period, sometimes by an order of magnitude.

\section{Astrophysical implications}
\label{implications}

	This analysis of the destruction timescale for metallicity inhomogeneities has strong astrophysical implications. First, we shown that a non-axisymmetric pattern in the metallicity field is smoothed in less than a galactic orbital period. This implies that non-axisymmetries in the metal field driven by spiral patterns, bar, or similar phenomena in the star formation distribution will be suppressed very rapidly. The underlying physical mechanism driving this is that illustrated in Figure \ref{mecha} and discussed by \citet{Yang2012}: for any non-axisymmetric mode, differential rotation winds the metal pattern up into a tight spiral on orbital timescales, and this small-scale pattern is then easily destroyed by turbulence. This mechanism likely explains why observed galaxy metallicity variations \red{in the gas} at fixed galactocentric radius are so small.
	
	However, the axisymmetric patterns seems more stable, particularly for $n\leq 2$ where $\tau_{n0}$ is bigger than $t_{\rm orb}$\footnote{$\tau_{10}$ is at least bigger than $\tau_{20}$ and probably bigger than several galactic orbital periods.}.  The much larger diffusion timescale for the axisymmetric modes likely explains why radial metallicity gradients in galaxies persist even as azimuthal ones are wiped out. Moreover, the diffusion time is also related to the timescale required for the metal distribution in a galaxy to reach equilibrium between star formation, which drives the metal distribution away from homogeneity, and turbulent mixing, which drives it toward homogeneity. Since the low $n$ axisymmetric modes may dominate the overall metallicity distribution, the relaxation time required for the metallicity to reach a steady state would be bigger than several orbits. This might explain why metal gradients for high redshift galaxies are far more varied than those in nearby galaxies. Most $z\simeq 0$ galaxies, including  the Milky Way, experienced their last major merger between $z=1-2$, and thus have been in their present configuration for $\sim 5-10$ Gyr. This is a timescale much larger than their orbital period, suggesting that most present-day galaxies have had time to reach equilibrium between metal production and diffusion. On the other hand, high redshift galaxies are often at most a few orbital periods old and are therefore still in a non-equilibrium state. As a result, their metal gradients have not undergone significant smoothing, and instead reflect the patchy distributions of star formation within them without much smoothing.

		\red{Finally, we end this discussion with a caution. We have argued that, because at present only isolated galaxy simulations can reach the resolutions requires to capture the multi-phase ISM, such simulations are the only way to derive realistic rates of metal transport. However, the limitation of this approach is that we have run a closed box simulation without gas infall or external perturbations. In a more realistic cosmological environment, the quantitative mixing timescales we have derived might be modified by the processes we have been forced to omit. However the differences between axisymmetric and non-axisymmetric should be preserved even in a cosmological context, since the main effect is purely geometrical.}

\section{Conclusions}
\label{conclusion}

	In this work we simulate the diffusion of inhomogenous metal distributions in a galactic disc as a result of gravitational instability-driven turbulence. To make our study as general as possible, we note that, in cylindrical geometry, the metal field can always be decomposed into Fourier-Bessel functions. We therefore study how different Fourier-Bessel modes decay and mix as a result of shear and turbulence. \red{We find that the efficiency of mixing strongly depends} on whether one is considering an axisymmetric or a non-axisymmetric inhomogeneity. In the \red{former} case, the metal field is very stable, destruction of the original pattern requires at least several orbital periods for large-scale modes, and is caused only by turbulence. In the \red{latter} case, the original pattern vanishes in less than a galactic orbital period. This difference in timescale is due to the effects of shear. Shear accelerates  diffusion by  winding up the inhomogeneities into tight spirals on small spatial scales, effectively transporting power from large to small scales in an orbital period. Once this transport is complete, turbulence is then able to diffuse the small-scale power quite rapidly.
	
	This difference between modes in terms of dissipation time has strong implications for our understanding of the observed distributions of metals in galactic discs. In particular, the far greater rapidity of mixing for non-axisymmetric modes than for axisymmetric ones helps explain why galaxies show consistent radial gradients in metallicity but little to no variation at fixed radius. The long mixing timescales we find for radial modes also suggest that metal distributions in high-redshift galaxies are most likely not yet in equilibrium between metal production and mixing. This provides a likely explanation for the much greater diversity of metallicity distributions seen at high redshift: these are more reflective of the patchy distribution of star formation in these galaxies, while the comparatively uniform behavior of low-$z$ galaxies arises from the balance between metal production and diffusion. Indeed the most important modes are not only those with the longest dissipation timescales, but also the most fed ones by stellar metal production. We leave consideration of that problem to future work.
	
\section*{Acknowledgements} 
This work was supported by NSF grants AST-0955300 and AST-1405962, NASA ATP grant NNX13AB84G, and NASA TCAN grant NNX14AB52G (MRK, NJG, and JCF).  The simulations for this research were carried out on the UCSC supercomputer Hyades, which is supported by the NSF (award number AST-1229745).
	
\appendix
\section{Convergence}

\begin{figure}
			\begin{center}
				\includegraphics[width=0.5 \textwidth]{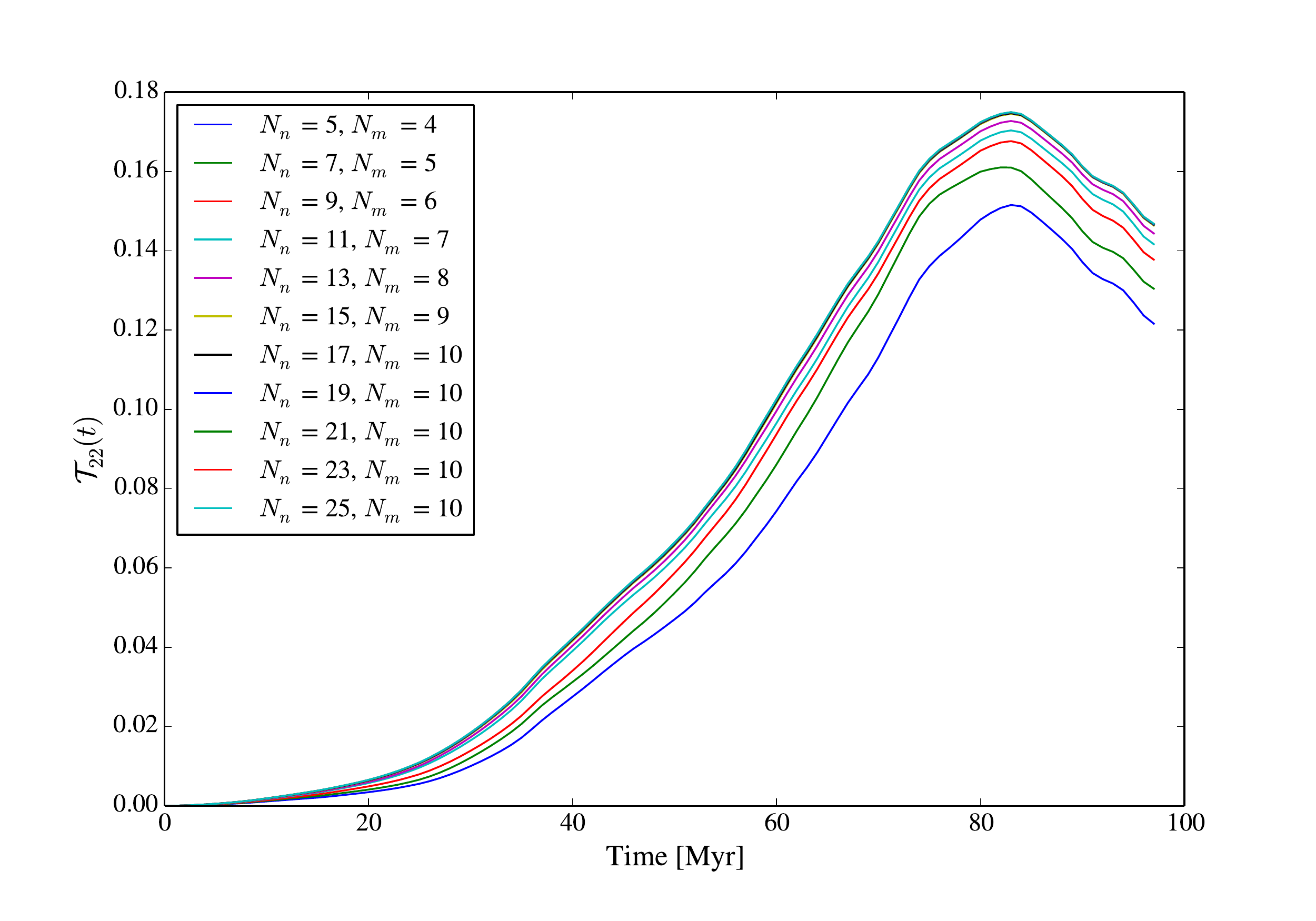}
				
				\caption{$\mathcal{T}_{22}$, computed using a sum truncated at the indicated values of $n'$ and $m'$.}
				\label{convergence}
			\end{center}
		\end{figure}

	For computational reasons, we must truncate the Fourier-Bessel expansions we use in our analysis at finite values of $n'$ and $m'$. We select as our limits $n'=25$ and $m'=10$, and in this Appendix we show that these limits are high enough to ensure convergence when we reconstruct fields such as $\chi^{\mathrm{sh}}$ and quantities like $\mathcal{T}_{nm}$ that depend on them. In Figure \ref{convergence}, we plot the quantity $\mathcal{T}_{22}$ computed using expansions truncated at various values for $n$ and $m$. As shown in the Figure, this quantity is very well converged by the time we reach $n'=25$, $m'=10$.
	
Another mean to check the convergence is to compute the difference between the norm $||\chi_{nm}||^2_2$ and the sum over the first terms of $\P_{nm,n'm'}^2$. Indeed, we have the equality:

\begin{equation}
	\int\chi^2\d S = \sum_{m=0,n=1}^\infty a_{nm}^2||J^c_{nm}||^2 + b_{nm}^2||J^s_{nm}||^2.
\end{equation}

We compute the ratio of the sum over $n=25$ and $m=10$ and $||\chi||^2$ for the 25 color fields studied after \mixtime . It appears that the fields lost less than 2\% for $n=1$, $m=0$, and less than 20\% for the highest $n$ and $m$ studied.

\bibliographystyle{mn2e}
\bibliography{library}

\label{lastpage}
\end{document}